# Faraday isolator based on crystalline silicon (c-Si) for 2 micron laser radiation

Ilya Snetkov,[1,*] and Alexey Yakovlev[1]

[1]*Federal Research Center Institute of Applied Physics of the Russian Academy of Sciences, 46 Ulyanov str, 603950,*

*Nizhny Novgorod, Russia*

[*]*e-mail:* [snetkov@appl.sci-nnov.ru](snetkov@appl.sci-nnov.ru)

The magneto-optical properties of single-crystal silicon were investigated as a function of wavelength and temperature. A bulk free-space traditional Faraday isolator for the radiation with a wavelength of 1940 nm with maximum admissible power $P_{max}$=20 W (magnetic field ~2.8 T) was implemented. The negative value of the piezo-optical anisotropy ratio of the used material allowed developing an Faraday isolator with compensation of thermally induced depolarization without reciprocal rotator with estimated $P_{max}$=180 W. The potential of single-crystal silicon as a magneto-optical material for Faraday isolators operating at room as well as at cryogenic temperature in a high-power laser radiation was considered. It was shown that single-crystal silicon is highly promising for the development of Faraday devices, including ones for next-generation laser interferometers aimed at detecting gravitational waves.

The devices based on the Faraday effect, such as Faraday isolators (FI) and rotators are the essential components of complex laser systems. Faraday isolators reduce the risk of amplifier self-excitation, protect master oscillator against unwanted feedback, and prevent damage of sensitive elements by back reflections. Faraday rotators are used to organize multipass amplifier schemes and schemes with compensation of thermally induced depolarization. Recently, there has been an increased interest in high-power laser radiation in the 1.4–3 µm range. Laser radiation in this range is broadly used in industry (welding, cutting, and drilling of plastics), medicine (surgery, urology), in research (material studies, atmospheric probing, environmental monitoring), and is promising for implementing submarine fiber-optic communication.[1] Also, the ~2 µm wavelength range is regarded to be highly promising for the new generation laser interferometers for gravitational wave detection.[2,3]

Whereas FIs providing an isolation ratio of better than 30 dB up to an average power of several kW (clear aperture Ø 30 mm) have been demonstrated for the radiation in the region of 1 µm,[4,5] only a few FIs ensuring high isolation at high average power have been reported in the wavelength range of ~2 µm (official web sites: EOT <30 W, clear aperture Ø4 mm; NewPort <50 W, clear aperture Ø4 mm). This is explained by both a lower availability of radiation sources with a high average power, and a smaller number of known and well-studied magneto-optical materials that are transparent in this wavelength range.

High-power laser radiation greatly affects the FI characteristics, thereby significantly narrowing the choice of magneto-optical materials. Therefore, an ideal magneto-optical material should possess the following properties: 1) low absorption at the operating wavelength; 2) high Verdet constant for maximal reduction of its length; 3) high mechanical strength to provide resistance to thermal stresses; 4) high thermal conductivity for fast heat removal and smoothing of temperature gradients; 5) a possibility to grow big bulk and produce a large aperture elements for operation with multimode radiation and for reducing risks of formation of nonlinear processes; and 6) to have thermo-optical characteristics guaranteeing a lower thermally induced birefringence at equal heat generation. One of the materials traditionally used for manufacturing magneto-optical elements (MOEs) of Faraday devices in this wavelength range is pure or doped yttrium-ferrum garnet REE:$Y_3Fe_5O_{12}$ (REE:YIG, REE=Bi, Ce, Dy, La and others) in single-crystal[6] or ceramic[7] form ($\Phi_{1940}$=221 rad/m)[8]. By virtue of its specific Faraday rotation in the saturated regime and a small value of magnetic field saturation, this ferrimagnetic found wide application in integrated optoelectronic devices.[9] However, it is rather hard to grow this material in the form of a large-size (>10 mm) single crystal of a high quality. Both ferrimagnetic single-crystals and ceramics have a relatively high absorption coefficient and due to low magnetic field saturation there arise difficulties in tuning the angle of Faraday rotation. Another material of interest is a diluted magnetic semiconductor CdMnTe [10] (V=35 rad/(T·m) [11]). However, it also has a relatively high absorption coefficient and difficulties in growing large-aperture samples of high optical quality. Other potential candidates are various chalcogenide glasses,[12] but they usually have a small thermal conductivity coefficient and cannot be used in laser radiation with high average power. Alternative prospective magneto-optical materials devoid of all or some of the above mentioned drawbacks are sesquioxide ceramics $Dy_2O_3$ ($V_{1940}$=13.8 rad/(T·m);[13] $V_{1940}$=25.9 rad/(T·m) [8,14]), $Er_2O_3$ ($V_{1940}$=6.2 rad/(T·m)))[15], $Yb_2O_3$ ($V_{1940}$=5.4 rad/(T·m))[16], fluoride single crystals $CeF_3$ ($V_{1940}$=8 rad/(T·m))[17], selenide ceramics ZnSe ($V_{1940}$=7.7 rad/(T·m))[18] and sulfides ZnS ($V_{1940}$=4.6 rad/(T·m))[19]. Some of these materials were used in Faraday isolators for high-power laser radiation.

A laser interferometer for gravitational wave detection is one of the unique devices providing new information about deep space and the interaction of massive astronomical objects.[20] The detection range and, hence, the amount of registered collisions or the sensitivity of registering less massive objects is fully determined by noises in the laser interferometer. To increase the sensitivity of the next-generation laser interferometers it is intended to use radiation cooling of test masses (for reduction of thermal noise), change the test-mass material from fused quartz to single-crystal silicon c-Si (for reduction of thermal noise), and use squeezed vacuum with a frequency dependent squeezing quadrature (for reduction of quantum noise), which will increase the sensitivity by a factor of ~10.[2] The choice of the c-Si material is based on the unusual behavior of its linear expansion coefficient on cooling: it becomes zero at a temperature of 123 K and 18 K.[21] In addition, c-Si has m3m cubic symmetry, a high thermal conductivity coefficient (150 W/(m·K) at 300 K and 600 W/(m·K) at 120 K)[22], can have a low absorption coefficient ($4.3·10^{-6}$ 1/cm (@1550 nm)),[23] and can be manufactured with a high-optical-quality aperture up to 450 mm without dislocations. The low absorption of the material from which the transmission optics is made is critical for maintaining the squeezed state, since any losses destroy the squeezed state and can annul its use for reducing quantum noise. However, as the transmission spectrum of c-Si is within 1.1–6.5 µm, 1 µm laser radiation cannot be used; hence, a shift to a longer wavelength region is demanded. The closest and well mastered region is near 2 µm, in which the emission lines of $Tm^{3+}$ and $Ho^{3+}$ ions located. Consequently, a Faraday isolator for this wavelength will be required.

The comparison of the requirements to magneto-optical material for FIs to be used in high-power radiation and of the properties of c-Si considered above shows that silicon is a good candidate in many respects; only the magneto-optical and thermo-optical properties of this material remain uncovered. The work described below is devoted to the study of magneto-optical properties of single-crystal silicon, the development of an FI on its basis, and the investigation of its isolation ratio during operation in high-power laser radiation.

The experiments were carried out using a high-purity c-Si material provided for us within the LIGO collaboration. First of all we investigated the wavelength dependence of the Verdet constant. For this we made use of the magnetic system of Nd-Fe-B magnets.[24] The general view and the magnetic field distribution along the axis of the system are presented in Fig. 1(a,b). Two cylindrical elements with L=12 mm and ø=11 mm (specified by the inner diameter of the magnetic system) with [001] orientation were made from the provided material (see Fig. 1(c)).

Laser diodes with 1310 nm and 1550 nm wavelengths and a Tm-fiber laser with a wavelength of 1940 nm were used as radiation sources. For wavelengths less than ~1100 nm the studied sample was opaque. The magnetic system was placed between crossed polarizers. Changes in the laser radiation intensity were controlled by the PyroCam IV (Ophir-Spiricon). By the angle of rotation of the output polarizer, at which the laser radiation intensity is minimized when the sample is placed in a magnetic field, we determined the angle of Faraday rotation and calculated the Verdet constant for each used wavelength. The values of the Verdet constant presented in Fig. 2(a) were $V_{1310}$=34.9 rad/(T·m), $V_{1550}$=23.8 rad/(T·m), and $V_{1940}$=15.2 rad/(T·m).

The values of Verdet constant as a function of wavelength measured in c-Si samples having a thickness of several tens of micrometers[25] are also shown in Fig. 2(a) for comparison. Our results obtained in a 12-mm thick sample agree well with the data known from the literature. Silicon is a diamagnetic material, so the experimental results were approximated in the form:[26]

$$V(\lambda) = \frac{1}{\lambda} \frac{n_0^2 - 1}{2n_0} \left[ A + \frac{B}{\lambda^2 - \lambda_0^2} \right], \qquad (1)$$

The approximation coefficients giving maximal coefficient of determination in the 1.2-3 μm region were $A$=7.9 μm·rad/(T·m); $B$=19.3 μm$^3$·rad/(T·m); and $\lambda_0$=1·10$^{-3}$ μm. The dispersion dependence of the index of refraction $n_0(\lambda[\mu m])=3.413+0.15/(\lambda^2-0.004^2)$ was obtained from the data from the Ref. [27] for room temperature.

We also investigated the behavior of the Verdet constant at cryogenic cooling in the scheme analogous to that used in our Ref. [28]. A sample of c-Si was placed in a copper sleeve and then into a cryostat cooled by liquid nitrogen. The magnetic system was installed outside the cryostat to avoid its cooling. The cryostat and the magnetic system were located between two polarizers. By the rotation of the second polarizer we determined the angle of Faraday rotation. The result of the experiment is shown in Fig. 2(b). To experimental accuracy, the Verdet constant in the studied temperature interval are temperature independent ($1/V \cdot dV/dT$=0). Consider the advantages of this fact: 1) a higher stability of faraday rotation angle with the variation of the ambient temperature, coolant, and laser radiation power. Significantly weaker requirements for thermal stabilization of the magneto-optical element and no need to use active thermal stabilization[29] and complicated mechanical design to tune position of MOE[30]; 2) no need to make a reserve of the MOE length so as to foresee a decrease in the angle of rotation with an increase in the average temperature of the element with increasing laser power[4]; 3) a higher efficiency of using [C] orientation,[31] FI schemes with compensation of thermally induced depolarization,[32] and a lower depolarization level in thin MOEs at cryogenic cooling[33] due to the absence of contribution to thermally induced depolarization from $\gamma_V$ which arises from the temperature dependence of the Verdet constant.

On cryogenic cooling of c-Si, the width of the fundamental bandgap changes, resulting in reduced absorption in the short-wavelength region and the respective shift of the absorption edge of the material to the short-wavelength region.[34] Thus, the thick sample at a wavelength of 1064 nm that was initially opaque at room temperature, became "transparent" for radiation at a temperature of 90 K, which allowed us to measure the Verdet constant $V_{1064}(T$=90 K$)$=45.7 rad/(T·m). The high value of thermal conductivity, the zero value of linear expansion coefficient (no expansion => no temperature stresses and no thermally induced birefringence), and transparency of the material open up opportunities for developing a cryogenic FI not only for lasers based on Tm and Ho (1900-2100 nm) ions, but possibly for lasers based on Nd and Yb (1053-1075 nm) ions as well, which is the subject matter of our further research.

The next step was a creation of a FI based on two cylindrical samples. In each of them, the directions of the crystallographic axes in the plane of the element end were determined with XRD (Bruker D8 Discover). The ends of the elements had AR dielectric coatings for the wavelength of 1940 nm. The linear absorption coefficient of the samples at 1940 nm was measured to be $α_0$=0.025 1/cm. Each sample was glued into a copper sleeve to provide heat sink. The sleeves were placed one after the other into the magnetic system, where their position in the magnetic field was aligned so that the angle of Faraday rotation in each element should be equal to 22.5º (which corresponds to the total angle of 45º), and the crystallographic axes directions in the samples were made coincide $θ_1=θ_2$ (imitation of a single MOE, $θ_i$ is the angle between the polarization of the radiation incident on the FI and one of the crystallographic axes). Further, we investigated the dependence of the thermally induced depolarization $γ=P_d/P_{laser}$ ($P_d$ – power of depolarized radiation) on the laser radiation power $P_{laser}$. A Tm-fiber laser with unpolarized radiation and a maximum laser power of 20 W was used as a source of laser radiation (10 W maximum power of polarized radiation). The output polarizer was adjusted to transmit only the depolarized field component. By simultaneous rotation, both elements were adjusted so as to observe minimal $γ$. In this position we measured $γ(P_{laser})$ (Fig. 3, red circles). The thermally induced depolarization demonstrated a characteristic behavior proportional to laser power squared. The estimated maximum permissible power of using a Faraday device $P_{max}$, at which the isolation ratio was $I=-10·\log_{10}(γ)$=30 dB, was ~20 W. The angle of Faraday rotation did not depend on $P_{laser}$ and was 45º throughout the studied range.

From the earlier works it is known that c-Si can have a negative piezo-optical anisotropy ratio $ξ$=-0.63±0.05.[35] We used the same silicon for producing MOEs. A specific feature of materials with $ξ$<0 is the presence of special orientations [C] and [P] in which thermally induced depolarization and thermal lens astigmatism caused by thermally induced birefringence[36,37] can be eliminated or reduced. These orientations exist, even in presence of Faraday rotation.[31,37] In addition, with the use of such materials, schemes with compensation of thermally induced depolarization can be implemented without reciprocal polarization rotator (quartz rotator),[38] including an FI.[32,39] So, we have developed an FI with compensation of thermally induced depolarization in the magnetic field without a quartz rotator. In this case, the elements rotated independent of each other $θ_1≠θ_2$, and at $θ_1≈$-2º and $θ_2≈$-40º the depolarization was compensated (Fig.3 a, blue squares). The direction of polarization rotation at Faraday rotation was taken as the positive change of the angle. The estimated values of the angles $θ_1$ and $θ_2$ for optimal compensation for the angles of Faraday rotation of 22.5º in each MOE made of material with $ξ$=-0.63 were $θ_1≈$-1.3º and $θ_2≈$-40.6º. The measured contrast of the scheme is shown by the black triangles in Fig. 3a.

As can be seen from the Fig. 3a, at the incidence power of 8.3, the isolation ratio is 38 dB and 42 dB for traditional FI and for FI with compensation of a thermally-induced depolarization, respectfully. The calculated curves for the traditional FI with a MOE in [001] orientation, for the FI with compensation of thermally induced depolarization in magnetic field using a MOE with [001] orientation without quartz rotator (like in the Ref. [39]), for the traditional FI with [C] orientation,[31] and for the FI with compensation in magnetic field using a MOE with [C] orientation[32] are presented in Fig. 3. The calculations were made using the linear absorption coefficient coinciding with the measured value of 0.025 1/cm. The calculations show that at $P_{max}$=20 W in the traditional FI, the use of the scheme with compensation allows increasing $P_{max}$ up to 180 W, the use of the [C] orientation up to $P_{max}$=500 W, and the simultaneous use of the scheme with compensation and crystals in [C] orientation up to 2 kW.

Let us estimate the potential of c-Si as a magneto-optical material. The samples of c-Si used in the FI had a relatively high linear absorption coefficient. High-purity, high-resistivity c-Si can have a low absorption coefficient $\alpha_0$=4.3·$10^{-6}$ 1/cm (@1550 nm)[23], which is more than 3 orders of magnitude lower than that in the studied crystal. The value of $P_{max}$ is inversely proportional to the value of the coefficient $\alpha_0$ and may be increased substantially with the use of a material of higher purity. From the point of view of the available data, c-Si has a huge potential as a magneto-optical material for Faraday isolators and rotators operating in the 2 µm high-power radiation. An additional advantage of using c-Si as magneto-optical for laser interferometers aimed at gravitational wave detection is that only one material can be employed in the facility. If all requirements for the optical quality and the value of linear absorption coefficient (for both the main radiation and the added squeezed light) of the c-Si material for test masses (55 cm thick) are met, this material may be readily used as a MOE (thickness < 4 cm for rotation by 45°). Another advantage is that c-Si is a non-toxic material unlike the other magneto-optical materials transparent in the 2µm range, such as ZnSe, $Te_2As_3Se_5$ or CdMnTe.

To conclude, we have investigated the wavelength and temperature dependence of the Verdet constant for c-Si. Bulk free-space FIs for the wavelength of 1940 nm according to two optical schemes – traditional and with compensation inside the magnetic field – based on MOE of a single-crystal silicon with [001] orientation have been implemented for the first time. The behavior of the isolation ratio and the dependence of the angle of Faraday rotation in the FI on the laser radiation power have been studied. The developed FIs ensure the isolation ratio of 38 dB and 42 dB at the incident radiation power of 8.3 W at the wavelength 1940 nm. The calculations demonstrate that the developed FIs will provide an isolation ratio better than 30 dB up to powers of 20 W and 180 W, respectively. Prospects of using c-Si as a magneto-optical material have been analyzed to show that the material has a very high potential for FIs and cryogenic FIs operating in high-power laser radiation.

**Acknowledgements**

The experimental part of reported study was funded by the RFBR (project number 19-2911019) and numerical calculation part of reported study was supported by the State Research Task for the Institute of Applied Physics, Russian Academy of Sciences (project number 0030-2021-0029). This paper was reviewed by Rodica Martin, the LIGO Scientific Collaboration, under LIGO Document P2100395.**Data availability**

The data that support the findings of this study are available from the corresponding author upon reasonable request.

**Figure captions**

**Fig. 1** a) General view of the FI's magnetic system; b) magnetic field distribution along the axis of the magnetic system; c) general view of the c-Si elements.

**Fig.2** a) Verdet constant *versus* wavelength for c-Si and Verdet constant values for some promising materials at 1940 nm wavelength; b) temperature dependence of c-Si Verdet constant at a wavelength of 1940 nm.

**Fig. 3** a) Experimental dependences of integral thermally induced depolarization on laser power for two FI schemes b) calculated dependences of integral thermally induced depolarization on laser power for different optical schemes of FI.

FIG.1

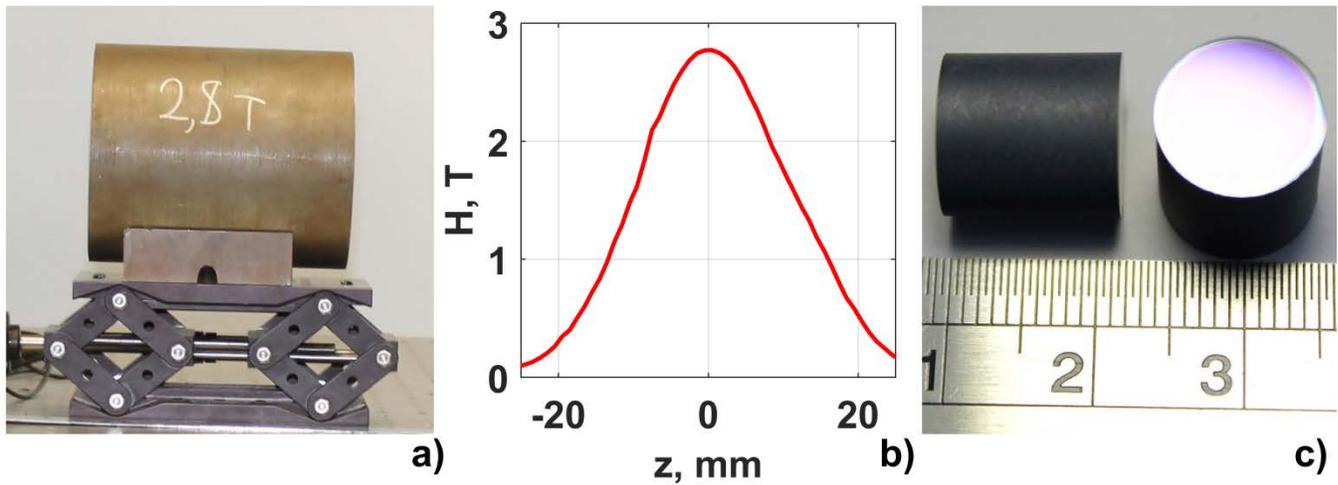

FIG.2

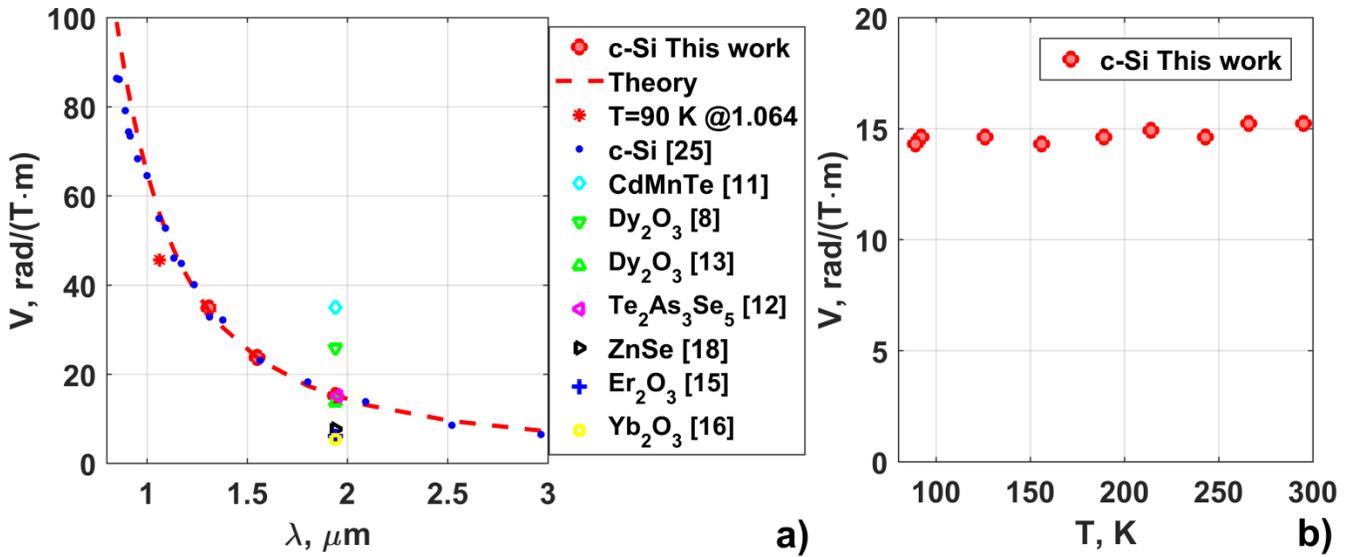

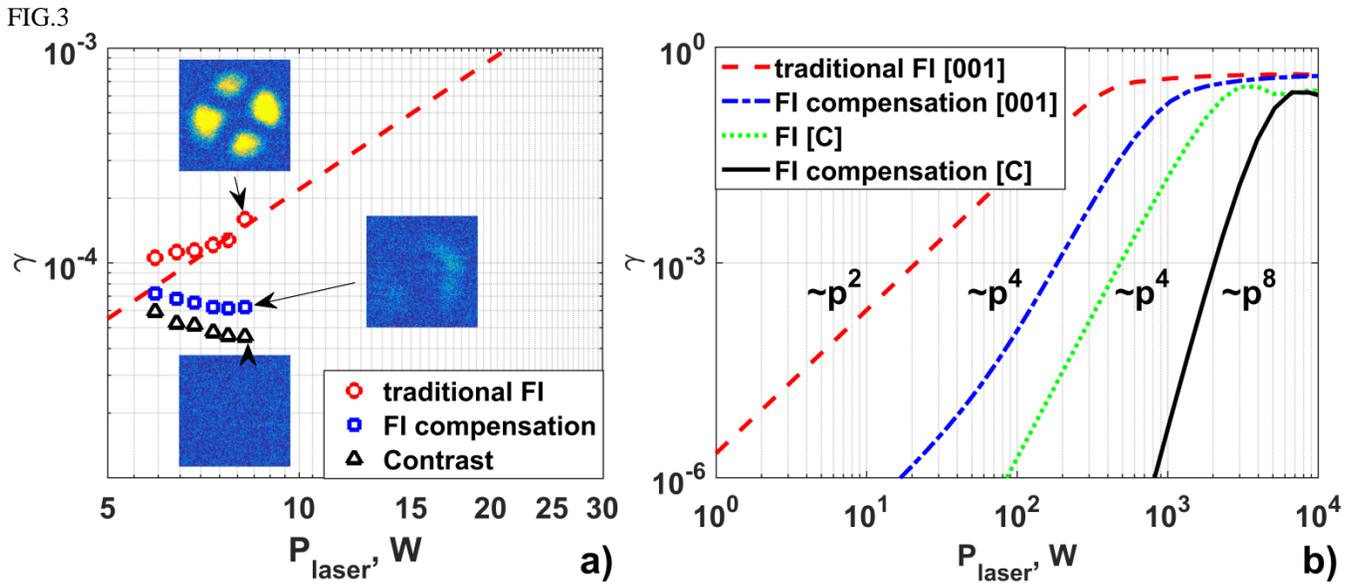

FIG.3

## References

1   Karsten Scholle, Samir Lamrini, Philipp Koopmann, and Peter Fuhrberg, in *Frontiers in Guided Wave Optics and Optoelectronics*, edited by Bishnu Pal (InTech, Rijeka, 2010), p. Ch. 21;   J. A. Savage,  Materials Science Reports **2** (3), 99 (1987).

2   R. X. Adhikari, K. Arai, A. F. Brooks, C. Wipf, O. Aguiar, P. Altin, B. Barr, L. Barsotti, R. Bassiri, A. Bell, G. Billingsley, R. Birney, D. Blair, E. Bonilla, J. Briggs, D. D. Brown, R. Byer, H. Cao, M. Constancio, S. Cooper, T.

Corbitt, D. Coyne, A. Cumming, E. Daw, R. deRosa, G. Eddolls, J. Eichholz, M. Evans, M. Fejer, E. C. Ferreira, A. Freise, V. V. Frolov, S. Gras, A. Green, H. Grote, E. Gustafson, E. D. Hall, G. Hammond, J. Harms, G. Harry, K. Haughian, D. Heinert, M. Heintze, F. Hellman, J. Hennig, M. Hennig, S. Hild, J. Hough, W. Johnson, B. Kamai, D. Kapasi, K. Komori, D. Koptsov, M. Korobko, W. Z. Korth, K. Kuns, B. Lantz, S. Leavey, F. Magana-Sandoval, G. Mansell, A. Markosyan, A. Markowitz, I. Martin, R. Martin, D. Martynov, D. E. McClelland, G. McGhee, T. McRae, J. Mills, V. Mitrofanov, M. Molina-Ruiz, C. Mow-Lowry, J. Munch, P. Murray, S. Ng, M. A. Okada, D. J. Ottaway, L. Prokhorov, V. Quetschke, S. Reid, D. Reitze, J. Richardson, R. Robie, I. Romero-Shaw, R. Route, S. Rowan, R. Schnabel, M. Schneewind, F. Seifert, D. Shaddock, B. Shapiro, D. Shoemaker, A. S. Silva, B. Slagmolen, J. Smith, N. Smith, J. Steinlechner, K. Strain, D. Taira, S. Tait, D. Tanner, Z. Tornasi, C. Torrie, M. Van Veggel, J. Vanheijningen, P. Veitch, A. Wade, G. Wallace, R. Ward, R. Weiss, P. Wessels, B. Willke, H. Yamamoto, M. J. Yap and C. Zhao, Classical Quant. Grav. **37** (16), 165003 (2020).

3    J. Eichholz, N. A Holland, V. B Adya, J. V van Heijningen, R. L Ward, B. J J Slagmolen, D. E McClelland, and D. J Ottaway, Physical Review D **102** (12), 122003 (2020).

4    I. L. Snetkov, A. V. Voitovich, O. V. Palashov, and E. A. Khazanov, IEEE J. Quantum Elect. **50** (6), 434 (2014).

5    E.A. Khazanov, Phys. Usp. **59** (9), 886 (2016); Evgeniy A. Mironov, Oleg V. Palashov, and Stanislav S. Balabanov, Opt. Lett. **46** (9), 2119 (2021).

6    Takenori Sekijima, Hiroshi Kishimoto, Takashi Fujii, Kikuo Wakino, and Masakatsu Okada, Jpn. J. Appl. Phys. **38** (Part 1, No. 10), 5874 (1999).

7    R. J. Young, T. B. Wu, and I. N. Lin, Mat. Res. Bull. **22** (11), 1475 (1987); Akio Ikesue and Yan Lin Aung, J. Am. Ceram. Soc. **0** (0), 1 (2018); Yimin Yang, Xiaoying Li, Ziyu Liu, Dianjun Hu, Xin Liu, Penghui Chen, Feng Tian, Danyang Zhu, Lixuan Zhang, and Jiang Li, Magnetochemistry **7** (5), 56 (2021).

8    David Vojna, Ondřej Slezák, Ryo Yasuhara, Hiroaki Furuse, Antonio Lucianetti, and Tomáš Mocek, Materials **13** (23), 5324 (2020).

9    Duanni Huang, Paolo Pintus, Chong Zhang, Paul Morton, Yuya Shoji, Tetsuya Mizumoto, and John E. Bowers, Optica **4** (1), 23 (2017).

10  D. U. Bartholomew, J. K. Furdyna, and A. K. Ramdas, Phys. Rev. B **34** (10), 6943 (1986); Y. Hwang and Y. Um, presented at the 2012 7th International Forum on Strategic Technology (IFOST), 2012 doi: 10.1109/ifost.2012.6357795.

11  Gary Stevens, Thomas Legg, and Peter Shardlow, Proc. of SPIE **9346**, 93460O (2015).

12  Masoud Mollaee, Pierre Lucas, Julien Ari, Xiushan Zhu, Michal Lukowski, Tariq Manzur, and N. Peyghambarian, Opt. Lett. **45** (8), 2183 (2020).

13  I. L. Snetkov, A. I. Yakovlev, D. A. Permin, S. S. Balabanov, and O. V. Palashov, Opt. Lett. **43** (16), 4041 (2018).

14  Y. L. Aung, A. Ikesue, R. Yasuhara, and Y. Iwamoto, Opt. Lett. **45** (16), 4615 (2020).

15  Alexey Yakovlev, Stanislav Balabanov, Dmitry Permin, Maxim Ivanov, and Ilya Snetkov, Opt. Mater. **101**, 109750 (2020).

16  Dmitry A. Permin, Anastasia V. Novikova, Vitaly A. Koshkin, Stanislav S. Balabanov, Ilya L. Snetkov, Oleg V. Palashov, and Ksenia E. Smetanina, Magnetochemistry **6** (4), 63 (2020).

17  Evgeniy A. Mironov, Aleksey V. Starobor, Ilya L. Snetkov, Oleg V. Palashov, Hiroaki Furuse, Shigeki Tokita, and Ryo Yasuhara, Opt. Mater. **69**, 196 (2017).

18  E. A. Mironov, O. V. Palashov, I. L. Snetkov, and S. S. Balabanov, Laser Physics Letters **17** (12), 125801 (2020).

19  M. Balkanski, E. Amzallag, and D. Langer, Journal of Physics and Chemistry of Solids **27** (2), 299 (1966).

20  LIGO Scientific Collaboration and Virgo Collaboration, Physical Review Letters **116** (6), 061102 (2016).

21  K. G. Lyon, G. L. Salinger, C. A. Swenson, and G. K. White, J. Appl. Phys. **48** (3), 865 (1977); Thomas Middelmann, Alexander Walkov, Guido Bartl, and René Schödel, Phys. Rev. B **92** (17), 174113 (2015).

22  T. Ruf, R.W. Henn, M. Asen-Palmer, E. Gmelin, M. Cardona, H.-J. Pohl, G.G. Devyatych, and P.G. Sennikov, Solid State Commun. **115** (5), 243 (2000).

23  Jerome Degallaix, Raffaele Flaminio, Danièle Forest, Massimo Granata, Christophe Michel, Laurent Pinard, Teddy Bertrand, and Gianpietro Cagnoli,  Opt. Lett. **38** (12), 2047 (2013).

24  E. A. Mironov, A. V. Voitovich, and O. V. Palashov,  Laser Physics Letters **17** (1), 015001 (2020).

25  Cedric J. Gabriel,  Phys. Rev. B **2** (6), 1812 (1970);  F. R. Kessler and J. Metzdorf,  physica status solidi (b) **60** (1), 125 (1973).

26  M. J. Weber, presented at the Laser and Nonlinear Optical Materials, San Diego, 1986 doi: 10.1117/12.939622.

27  H. H. Li,  Journal of Physical and Chemical Reference Data **9** (3), 561 (1980).

28  I. L. Snetkov and O. V. Palashov,  Opt. Mater. **62**, 697 (2016).

29  O.V. Palashov, I.B. Ievlev, E.A. Perevezentsev, E.V. Katin, and E.A. Khazanov,  Quantum Electron. **41** (9), 858 (2011).

30  Eric Genin, Maddalena Mantovani, Gabriel Pillant, Camilla De Rossi, Laurent Pinard, Christophe Michel, Matthieu Gosselin, and Julia Casanueva,  Appl. Opt. **57** (32), 9705 (2018).

31  I. L. Snetkov,  IEEE J. Quantum Elect. **54** (2), 1 (2018).

32  I. L. Snetkov,  IEEE J. Quantum Elect. **57** (5), 1 (2021).

33  D.S. Zheleznov, A.V. Starobor, O.V. Palashov, and E.A. Khazanov,  Journal of the Optical Society of America B **29** (4), 786 (2012).

34  G. G. Macfarlane, T. P. McLean, J. E. Quarrington, and V. Roberts,  Journal of Physics and Chemistry of Solids **8**, 388 (1959);  R. Pässler,  physica status solidi (b) **216** (2), 975 (1999).

35  I. Snetkov, D. Bulanov, A. Yakovlev, O. Palashov, and E. Khazanov,  Opt. Lett. (to be published doi:10.1364/OL.446337).

36  I.L. Snetkov, A.G. Vyatkin, O.V. Palashov, and E.A. Khazanov,  Opt. Express **20** (12), 13357 (2012).

37  A. Yakovlev and I. L. Snetkov, IEEE J. Quantum Elect. **56** (4), 6100108 (2020).

38  I.L. Snetkov and O. V. Palashov, Appl.Phys. B **109** (2), 239 (2012); Anton G. Vyatkin, Ilya L. Snetkov, Oleg V. Palashov, and Efim A. Khazanov, Opt. Express **21** (19), 22338 (2013); Ilya L. Snetkov, Vitaly V. Dorofeev, and Oleg V. Palashov, Opt. Lett. **41** (10), 2374 (2016).

39  A. V. Starobor, I. L. Snetkov, and O. V. Palashov, Opt. Lett. **43** (15), 3774 (2018).